\documentclass[pra,floatfix,twocolumn,showpacs,superscriptaddress]{revtex4}
\usepackage{amsmath}
\usepackage{latexsym}
\usepackage{bm}
\usepackage{amssymb}
\usepackage{graphics}
\usepackage{bbm}
\newcommand{\Tr}{\mathrm{Tr}}
\newcommand{\tr}{\mathrm{Tr}}
\newcommand{\fl}{\hspace{-1cm}}
\newcommand{\ket}[1]{|#1\rangle}
\newcommand{\bra}[1]{\langle #1|}
\newcommand{\?}{\mathrm{?}}

\begin{document}

\title[Single-experiment-detectable multipartite entanglement witness]{Single-experiment-detectable multipartite entanglement witness for ensemble quantum computing}

\author{Robabeh Rahimi}
\email{rahimi@math.kindai.ac.jp}
\affiliation{Department of Physics, Kinki Universtity, 3-4-1 Kowakae, Higashi Osaka, Osaka 577-8502, Japan}
\author{Akira SaiToh}
\email{saitoh@qc.ee.es.osaka-u.ac.jp}
\affiliation{Graduate School of Engineering Science, Osaka University, 1-3 Machikaneyama, Toyonaka, Osaka 560-8531, Japan}
\author{Mikio Nakahara}
\affiliation{Department of Physics, Kinki Universtity, 3-4-1 Kowakae, Higashi Osaka, Osaka 577-8502, Japan}
\author{Masahiro Kitagawa}
\affiliation{Graduate School of Engineering Science, Osaka University, 1-3 Machikaneyama, Toyonaka, Osaka 560-8531, Japan}

\begin{abstract}
In this paper we provide an operational method to detect multipartite entanglement in ensemble-based quantum computing. This method is based on the concept of the entanglement witness. We decompose the entanglement witness for each class of multipartite entanglement into nonlocal operations in addition to local measurements. Individual single- qubit measurements are performed simultaneously; hence complete detection of entanglement is performed in a single- run experiment. In this sense, our scheme is superior to the generally used entanglement witnesses that require a number of experiments and preparation of copies of quantum state for detection of entanglement.

\end{abstract}


\maketitle

\section{Introduction}
In view of the rapid development of the experimental realization of quantum information processing (QIP) by using different physical systems, it is an urgent obligation to find a proper strategy for detection of entanglement. One of the most commonly used systems in the study of physical realizations of QIP is an ensemble system and the best-known technology therein is nuclear magnetic resonance (NMR) \cite{NMR}.

The ensemble system of NMR has been employed for implementation of even relatively complicated quantum algorithms \cite{2,3,4}. However, NMR has been facing difficulty with regard to the existence of entanglement \cite{5,6}. The complication actually arises because of somewhat confusing ensemble behavior of the NMR system. The confusion manifests itself when macroscopic quantities, such as results of NMR measurement involving an average over a large number of molecules, are used to detect microscopic properties, such as entanglement between pairs of nuclear spins inside each molecule. In fact, it has been shown that, for some particular case of NMR implementation of quantum nonlocal algorithms, the apparent nonlocal behavior of the highly mixed states in NMR is due to a large number of molecules involved in the ensemble system \cite{self}. Hence, highly mixed states of NMR are separable and cannot be used for immaculate implementation of quantum nonlocal algorithms for which entanglement is believed to play an essential prerequisite role \cite{entim1, entim2}.

On the other hand, ensemble quantum computing (QC) pertains some inevitable advantages. It is particularly workable since spin manipulation is performed easily by applying corresponding pulses. In addition, ensemble QC, such as NMR, is supported by a long term research in the area of spectroscopy. Thus, it would be unfair and also inefficient to totally ignore ensemble QC. We have been involved in realization of QIP by means of electron nuclear double resonance spectroscopy (ENDOR) \cite{ourendor}. Although this system should be taken as an ensemble system at the moment, it has been profoundly evaluated for more elaborated nonlocal QIP through experimental studies \cite{mehring1, mehring2}. However, after applying entangling operations in an experiment, it might still be premature to claim that the state is properly entangled enough for implementation of nonlocal QIP. At least some qualitative detection of general multipartite entanglement for a particular system of interest should be examined in advance.

Detection of a general multipartite entanglement is one of the most challenging problems for an experimental study in QIP. There are several approaches in this context. One may first employ a full state tomography with which the complete density matrix would be obtained. Then, direct application of an existing entanglement measure would be evaluated on the quantum state in order to extract information about the entanglement of the state. This is, however, a bit too general to be efficient. The density matrix of the quantum state includes far more information than necessary for an entanglement estimation only. Furthermore, it is difficult to find a sufficient condition of entanglement for a given state in general if the dimension of the state is neither $2\times 2$ nor $2\times 3$ for which the Peres-Horodecki criterion \cite{peres1, peres2} would be applicable. Detection of entanglement of a quantum state through violation of the Bell inequalities also should not be taken to be perfect since there are entangled states that do not violate any known Bell inequalities \cite{bell}.

Still one may try the existing approaches that work for a state that is totally unknown in advance \cite{Sancho, Horedecki, ekert}. However, this may not be the most proper choice for our system of interest since as long as we are working in an experiment, we have definitely some knowledge about the state, i.e., through the prepared initial state or applied operations. If this is the case, it would be more appropriate to define an entanglement detector measure that is easily workable with less experimental effort by taking advantage of the available information on the state. We have studied the concept of entanglement witness (EW) \cite{terhal, lewen} with the motivation to introduce an entanglement detection applicable for a particular system of interest. The EW is an observable which has non-negative expectation values for all separable states. Therefore, detection of a negative value indicates the entanglement of the state.

In addition to the very fast development in the theory of EWs for different classes of states \cite{lewen, lewen2, brun, Toth2, GuhneO}, in experiments also detection of entanglement by the use of witness operators has attracted special attention \cite{Barn, Altn, Kien, Lein, Hafn, Boun}. For physical systems in thermodynamical limits, witness operators are developed \cite{Totn, Brunn, Down, Jorn, Wun, Totnn}. In addition, EWs are generated for detecting entanglement of mixed states that are close to a given pure state \cite{AcinPRL, Hyln}. EWs are used for characterizing different classes of a multipartite quantum state \cite{GuhneJMP50}. 

After determination of the EW for a particular expected state in an experiment, the important task is to decompose it into local operators that are easily measurable in the given physical system. This approach for detection of entanglement is operationally possible and, for most cases, is a simple method \cite{guhne, GuhneJMP50, GuneIJofTheorPHYS42}. In Refs.\ \cite{bourennanePRL922, Toth} detection of multipartite entanglement with few local measurements is studied. Throughout the study on this issue, it has been more appreciated to operationally simplify the entanglement detection process by modifying the required observables for the particular working system and/or decreasing the number of local measurements.

Generally speaking, previously introduced methods for detection of entanglement by the use of EWs require several projective measurements on the copies of a state in order to extract the outcome. However, it is somehow operationally difficult to prepare several copies of a quantum state to be measured for detection of entanglement. Therefore, it is more advantageous if an EW works just with a single run experiment. Also, the detection process would be better to be specially modified for the particular working system. For instance, if the physical system of interest is an ensemble system, then the available measurements are ensemble average measurements.

We work with ensemble systems. Therefore, ensemble average measurements should be employed here, in contrast with the projective measurements widely used in the previous works. Therefore, in this work, we propose a proper single-experiment-detectable (SED) EW for the ensemble system QC. The most significant result of our work is that our schemes require only a single-run experiment to detect entanglement by using nondestructive ensemble average measurements, which allow us to measure several non-commuting operators simultaneously. A well known example of nondestructive measurement is the free induction decay measurement that is frequently used in NMR quantum computing.

This paper is organized as follows. First in the following section we will briefly summarize the concept of the EW and will give a short review particularly for multipartite EWs. Then, in the third section we will introduce and prove a method with which the conventional EW can be transformed into a collection of separate but simultaneous measurements of individual local systems of the ensemble QC. The analysis of this section is based on the assumption that the density matrix is diagonal after application of the disentangling operations introduced there. In the subsequent section, a different SED EW, which works without any assumption on the density matrix, is introduced. As a drawback, however, we have to introduce an ancillary qubit. In Sec.\ \ref{five} we will extend discussions on our scheme, study the complexity of the corresponding quantum circuits for SED EW, and give some remarks on the behavior of this scheme under generally existing noise in the system.

\section{EW for operational detection of entanglement}
A density matrix $\rho$ is entangled if and only if there exists a Hermitian operator $W=W^\dagger$, called an entanglement witness (EW), such that
\begin{equation}
\label{define}
\left\{\begin{array}{l}
{\rm Tr} (W \rho) <0, \\
{\rm Tr} (W \sigma) \ge 0, \hspace{1cm} \forall \sigma \in S,
\end{array}\right.
\end{equation}
where $S$ denotes the set of separable states \cite{terhal2, cavalcanti}. Therefore, it is concluded that a state is entangled if some negative value is obtained in measurement of an EW.

There are several methods to construct an EW \cite{guhne}. For the case in which a density matrix has a negative eigenvalue when partially transposed, the construction of the EW is very simple. The partially transposed projector onto the eigenvector corresponding to the negative eigenvalue of the partial transpose of the state is an EW \cite{GuhneJMP50}.

One important and notable point about EWs appears when dealing with the multipartite case. EWs can be used to detect different kinds of multipartite entanglement by defining the space $S$ in Eq. (\ref{define}) to be the set of states that does not have a special kind of entanglement to be detected \cite{GuhneJMP50}. Let us give an example.  While all the entanglements for two qubits are actually equivalent to each other, for three qubits there are two classes of genuine pure tripartite entangled states \cite{Durn}. The entangled states from different classes cannot be transformed into each other by local operations and classical communications (LOCC). One class is the Greenberger-Horne-Zeilinger (GHZ) class that includes entangled states LOCC-equivalent to \cite{GuneIJofTheorPHYS42}
\begin{equation}
\label{ghz}
|{\rm GHZ}\rangle=\frac{1}{\sqrt 2}(|000\rangle+|111\rangle).
\end{equation}
The other tripartite class of entangled states is the Werner (W) class whose representative vector is
\begin{equation}
|{\rm W}\rangle=\frac{1}{\sqrt 3} (|100\rangle+|010\rangle+|001\rangle ).
\end{equation}
This classification can be extended to more general tripartite mixed states. We remind the reader that a tripartite (and generally any multipartite) state $\rho$ is separable if it can be written as a convex combination of fully separable states. However, a state $\rho$ may not be fully separable but biseparable, i.e., it can be written as a convex combination of biseparable pure states. If a tripartite state $\rho$ is neither fully separable nor biseparable then it is entangled, in either the GHZ class or the W class \cite{AcinPRL}. Corresponding to each class of entangled states, there are EWs already known. Needless to say, these entanglement witnesses are not sufficient to decide if the system possesses a genuine multipartite entanglement. For instance, for the GHZ state, the EW is 
\begin{equation}\label{convEWGHZ}
W_{\rm GHZ}= \frac{3}{4}{\mathbbm{1}}-|{\rm GHZ}\rangle\langle{\rm GHZ}|,
\end{equation}
while for the W state, it is 
\begin{equation}
W_{\rm W}= \frac{1}{4}{\mathbbm{1}}-|{\rm W}\rangle\langle{\rm W}|,
\end{equation}
where ${\mathbbm{1}}$ is the identity operator \cite{AcinPRL}.

The EW that detects a genuine tripartite entanglement of a pure state $|\Psi\rangle$, and of states close to $|\Psi\rangle$, is given by \cite{bourennanePRL922, Toth, Toth2}
\begin{equation}
\label{ew}
W= c{\mathbbm{1}}-|\Psi\rangle\langle\Psi|,
\end{equation}
where
\begin{equation}
\label{alpha}
c= \max_{|\Phi\rangle\in B} |\langle \Phi|\Psi\rangle|^2
\end{equation}
and $B$ denotes the set of biseparable states. Then $\Tr(W\rho_B)\ge 0$ for all biseparable states $\rho_B$ and ${\rm Tr}(W|\Psi\rangle\langle\Psi|)<0$. The coefficient $c$ in Eq. (\ref{alpha}) is determined by using the Schmidt decomposition \cite{bourennanePRL922}.

For an ensemble system, it is desirable to find an observable that can be performed by a small number of experiments to satisfy operational requirements. Any EW that is defined for the state $|\psi\rangle$, which is the most expected state after the applied operations and is close to the experimentally realized state $\rho$, should be decomposed into a linear combination of individual polarization operators, in order to make it locally measurable. In this paper, we introduce a new method with which an EW for multipartite states can be measured just with a single-run experiment.

\section{Single-experiment-detectable entanglement witness}
For starting up an experiment, the state is supposed to be initialized to a simple fiducial state, such as $|\psi\rangle=|0_1 ... 0_n \rangle$, for an $n$-qubit system. From now on  we will drop the subscripts $1, \ldots , n$ to simplify notation, unless otherwise stated. However, in an ensemble QC, the initial state is prepared in the form of a pseudopure state as follows
\begin{equation}
\label{pps}
\rho=(1-\epsilon){\mathbbm{1}}/2^n +\epsilon |\psi\rangle\langle\psi|,
\end{equation}
where $\epsilon$ characterizes the fraction of the state $|\psi\rangle$. It is important to note here that the confusion regarding the concept of entanglement in an ensemble system is intrinsically apart from the concept of pseudopure state as an initial state. A pseudopure state is used for making the required input state for QC.  Improvement of experimental conditions above some definite threshold is required to realize an entangled state experimentally. In other words, the pseudopure state still can be used for realization of a genuine entangled state if the experimentally required conditions are all satisfied. Here, we do not enter into the discussion of the very large required number of steps for making a pseudopure state as we suppose that, anyhow, we are given some prepared input state for which the status of entanglement should be studied.

In order to produce a particular entangled state, the corresponding entangling operation $V$ is applied on a state 
$|\psi\rangle$ to yield
\begin{equation}
|\psi\rangle_{\rm in}=V|\psi\rangle.
\end{equation}
The corresponding pseudopure state is
\begin{equation}
\label{pps}
\rho_{\rm in}=(1-\epsilon){\mathbbm{1}}/2^n +\epsilon |\psi\rangle_{\rm in}\, {}_{\rm in}\langle\psi|.
\end{equation}
Note that, even though the entangling operation $V$ is applied, $\rho_{\rm in}$ may or may not be entangled, and our task is to detect entanglement of $\rho_{\rm in}$ in order to examine whether $\rho_{\rm in}$ is applicable for some quantum nonlocal algorithm, for example.

The corresponding conventional EW, Eq. (\ref{ew}), for $|\psi\rangle_{\rm in}$ is \cite{bourennanePRL922, Toth, Toth2}
\begin{equation}
W_{\rm conv}=c {\mathbbm{1}}-|\psi\rangle_{\rm in} \, {}_{\rm in}\langle\psi|
            =c {\mathbbm{1}}-V|\psi\rangle\langle\psi|V^{\dagger},
\end{equation}
where $c=c(|\psi\rangle_{\rm in})$ is determined properly so that we do not get a negative value for separable states \cite{bourennanePRL922}. The entanglement witness $W_{\rm conv}$ detects $\rho_{\rm in}$ as entangled if $\epsilon > \epsilon_{\rm limit}$ where
\begin{equation}
\label{cond}
\epsilon_{\rm limit}:= \frac{{\rm Tr}(W_{\rm conv})}{{\rm Tr}(W_{\rm conv})-2^{n}\,{}_{\rm in}\langle\psi| W_{\rm conv} |\psi\rangle_{\rm in}}.
\end{equation}
For instance, if a state $\rho_{\rm GHZ}$ is the experimentally achieved pseudopure state for the state $|{\rm GHZ}\rangle$ Eq.~(\ref{ghz}), then $\epsilon_{\rm limit}$ is $5/7$. This large value of $\epsilon$ is in accordance with the results of studies on entanglement of pseudopure states which are indeed mixed states \cite{6}. One may still work with entanglement for an ensemble system without being engaged in the concept of a pseudopure state \cite{yu}. However, if the state should be used for a quantum computation it is a formidable task to prepare a pseudopure state that works exactly as a pure state under unitary operations for QC.

Generally speaking, the expectation value of the observable $W_{\rm conv}$ is given after several measurements on copies of state $\rho_{\rm in}$. If ${\rm Tr} (\rho_{\rm in} W_{\rm conv}) < 0$ then $\rho_{\rm in}$ is entangled. In this work, we introduce a strategy with which the EW can be detected in a single run-measurement. 

The most usual and convenient measurement in an ensemble QIP is the spin magnetization $z_i$ of the $i$th spin. Then, we find the proper $n$-partite EW, which we call $W$, that satisfies the following equation
\begin{equation}
\label{equality}
{\rm Tr} \left(\rho_{\rm in} W \right)={\rm Tr}\left(
 \rho_{\rm in} W_{\rm conv}\right),
\end{equation}
and
is detectable in a single run measurement as long as $W$ is written as
\begin{equation}
\label{new}
W=a_0+U^\dagger
\left(\sum_{k=1}^n a_k I^{\otimes n-k}\otimes Z\otimes I^{\otimes k-1}
\right) U.
\end{equation}
The unitary operator $U$ should be appropriately defined in addition to the coefficients $\{ a_i \}_{i=0}^{n}$. It should be noted that Eq.~(\ref{equality}) is state dependent.

\begin{figure}[hpt]
\begin{center}
\includegraphics{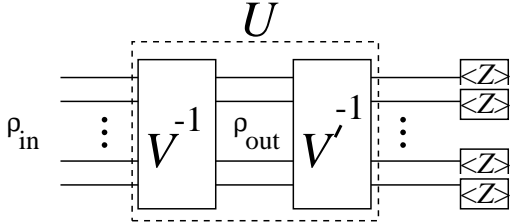}
\caption{\label{figone} It is possible to use a unitary transformation $U$ and individual polarizations of output qubits to find the value of 
$\Tr(\rho_{\rm in}W_{\rm conv})$. It is assumed that $\rho_{\rm out}$ is 
diagonal. See the text for details.}
\end{center}
\end{figure}

Now we will show that there exists some unitary operator $U$ in addition to the set of coefficients $\{ a_i \}_{i=0}^{n}$ such that Eq.~(\ref{equality}) is satisfied. We also give an explicit example of the solution. The unitary operator $U$ includes the inverse entangling operation $V^{-1}$ and $V^{{\prime}^{-1}}$ that is introduced for Eq.~(\ref{equality}) to be satisfied as in Fig.\ \ref{figone}.

Suppose $\rho_{\rm out}$, the state after applying the inverse entangling operation $V^{-1}$, is a diagonal matrix with a possible classical correlation. Although this assumption is often satisfied, it can be dropped for a general proof if an ancillary qubit is added to the quantum circuit. Indeed, for this case the only required measurement would be on the spin magnetization of the ancillary qubit. This is discussed in the next section.

In the present case, out task is to find some proper unitary transformation $U$ and coefficients $\{a_k\}_{k=0}^n$ such that Eq. (\ref{equality}) holds under the condition that $V^\dagger\rho_{\rm in}V$ is diagonal. We first prove by induction that Eq. (\ref{equality}) is satisfied for any $n$ and later give an explicit example of tripartite states. By setting $U={V'}^\dagger V^\dagger$ and $a_0=c+b$, we find for the left and right-hand sides (LHS and RHS) of Eq.~(\ref{equality}) 
\begin{eqnarray}\label{lhs}
{\rm (LHS)} 
&=&c-\Tr(\rho_{\rm out}\ket{0\ldots0}\bra{0\ldots0})\\
\label{rhs}
{\rm (RHS)}  
&=&c+\Tr \bigl[\rho_{\rm out}{V'}\bigl(b\notag\\
&{}&\hspace{0.7cm}+\sum_{k=1}^n a_k I^{\otimes n-k}\otimes Z\otimes I^{\otimes k-1}\bigr)
{V'}^\dagger\bigr].
\end{eqnarray}

Consider the case where $n=2$ and let us denote the matrix $V'$ by ${V'}_2$. By inspecting Eqs.~(\ref{lhs}) and (\ref{rhs}), we immediately notice that there is a solution in which the $(0 \ldots 0, 0 \ldots 0)$ element of ${V'}\left(b+\sum_{k=1}^n a_k I^{\otimes n-k}\otimes Z\otimes I^{\otimes k-1}\right) {V'}^\dagger$ is $-1$ while all the other diagonal elements vanish. Note that the off-diagonal elements are arbitrary thanks to the assumed diagonal form of $\rho_{\rm out}$. Typically, we can take $b=-1/4$, $a_1=a_2=3/8$, and
\begin{equation}
{V'}_2=\left(
\begin{array}{cccc}
0&0&0&1\\
\frac{1}{\sqrt{3}}&\frac{1}{\sqrt{3}}&
\frac{1}{\sqrt{3}}&0\\
\frac{1}{\sqrt{3}}e^{i2\pi/3}&\frac{1}{\sqrt{3}}e^{-i2\pi/3}&
\frac{1}{\sqrt{3}}&0\\
\frac{1}{\sqrt{3}}e^{-i2\pi/3}&\frac{1}{\sqrt{3}}e^{i2\pi/3}&
\frac{1}{\sqrt{3}}&0
\end{array}\right).
\end{equation}
Then,
\begin{eqnarray}
\label{last}
 {V'}_2(b&+&a_1 I\otimes Z+ a_2 Z\otimes I){V'}_2^\dagger \notag\\
 &{}&\hspace{-0.4cm}=
\left(\begin{array}{cccc}
-1&0&0&0\\
0&0&\frac{1}{4}e^{-i2\pi/3}&\frac{1}{4}e^{i2\pi/3}\\
0&\frac{1}{4}e^{i2\pi/3}&0&\frac{1}{4}e^{-i2\pi/3}\\
0&\frac{1}{4}e^{-i2\pi/3}&\frac{1}{4}e^{i2\pi/3}&0
\end{array}\right)
\end{eqnarray}
as promised. Therefore, the equality Eq. (\ref{equality}) is satisfied for the case $n=2$. We will take advantage of this typical ${V'}_2$ and the corresponding parameters in the following proof.

Now we use mathematical induction. Suppose that ${V'}_n$ transforms 
$$
b + a_1 I\otimes \ldots \otimes I\otimes Z + \ldots +a_n Z\otimes I\otimes \ldots\otimes I
$$
to ${\rm  diag}(-1_1,0_2,\ldots,0_{2^n}) + \mbox{off-diagonal terms}$. Then, we show that with a class of parameters, the unitary transformation ${V'}_{n+1}=U_{{\rm  bd}n+1}U_{{\rm  p}n+1}(I\otimes {V'}_n)$ maps
\begin{eqnarray*}
\frac{1}{2}(b + a_1 I\otimes \ldots \otimes I\otimes Z + \ldots &+& a_n I\otimes Z\otimes I\otimes \ldots\otimes I)\\
 &+& a_{n+1}Z\otimes I\otimes\ldots\otimes I
\end{eqnarray*}
to ${\rm  diag}(-1_1,0_2,\ldots,0_{2^{n+1}}) + \mbox{off-diagonal terms}$. The unitary operator $U_{{\rm  p}n+1}$ is a permutation operation that comprises transpositions: $(2,2^n+1)$, $(4,2^n+3)$, $\ldots$, $(2^n,2^{n+1}-1)$. This set of transpositions may be written in terms of ket vectors in the binary system: 
$(\ket{0_10_2\ldots0_n1_{n+1}},\ket{1_10_2\ldots0_n 0_{n+1}})$, 
$(\ket{0_10_2\ldots0_{n-1}1_n1_{n+1}},\ket{1_10_2\ldots0_{n-1}1_n0_{n+1}})$, 
$\ldots$, $(\ket{0_11_2\ldots1_n1_{n+1}},\ket{1_11_2\ldots1_n0_{n+1}})$,
or equivalently  
\begin{equation}
U_{{\rm p}n+1} = \sum_{i_k \in \{0, 1\}}
|i_{n+1} i_n \ldots i_2 i_1 \rangle \langle i_{1}i_n \ldots i_2 i_{n+1}|.
\end{equation}
Thus this is nothing but a SWAP operation between the first qubit and the $(n+1)$th qubit. We also defined $U_{{\rm  bd}n+1}={\rm  diag}(I,H,\ldots,H)$. This unitary transformation is easy to implement. By noting the equation $U_{{\rm bd}n+1}=(I^{\otimes n}\otimes H)~{\rm  diag} (H,I,\ldots I)$ we find that a single Hadamard gate acting on the $(n+1)$th qubit and a single C$^n H$ gate with zero-conditional-control qubits $1,\ldots,n$ and the target qubit $n+1$ do the job.

Now we give the details of the proof. Suppose that for $n$ qubits we have the unitary operator ${V'}_n$ and coefficients $b$ and $\{a_i\}_{i=1}^n$ that satisfy Eq.~(\ref{equality}) so that
\begin{eqnarray}
A_n&=&{V'}_n(b+ a_1 I\otimes\ldots\otimes I\otimes Z \notag\\
&{}&\hspace{0.9cm}+ \ldots+a_n Z\otimes I \otimes\ldots\otimes I){V'}_n^\dagger\nonumber\\
&=&
\left(\begin{array}{ccccc}
-1&\?&\?&\ldots&\?\\
\?& 0&\?&\ldots&\?\\
\vdots& &\ddots& &\?\\
\?& \?& \ldots & 0 &\?\\
\?& \?& \ldots & \? &0
\end{array}\right).
\end{eqnarray}
Here also the off-diagonal terms are not important. Then, for $n+1$ qubits, we have the equation
\begin{eqnarray}
\label{bn1}
B_{n+1}&=&I\otimes {V'}_n\Big[
\frac{1}{2}(b+ a_1 I\otimes\ldots\otimes I\otimes Z \notag\\
&{}&\hspace{2.0cm}+ \ldots+a_n I\otimes Z\otimes I \otimes\ldots\otimes I)\nonumber\\
&{}&\hspace{1.4cm}+a_{n+1}Z\otimes I\otimes\ldots\otimes I
\Big]I\otimes{V'}_n^\dagger\nonumber\\
&=&\frac{1}{2}
\left(\begin{array}{cc}
A_n&0\\
0&A_n
\end{array}\right)\\
&{}&+a_{n+1}\; {\rm  diag}(1_1,\ldots,1_{2^n},-1_{2^n+1},\ldots,-1_{2^{n+1}}).\notag
\end{eqnarray}
This should be transformed to the form: ${\rm  diag}(-1,0,\ldots,0)+ \mbox{off-diagonal terms}$ in order for Eq. (\ref{equality}) to be satisfied. To this end, we apply some unitary transformations to $B_{n+1}$. First, we use the permutation $U_{{\rm  p}n+1}$. Then we have 

\begin{eqnarray}
&&\hspace{-0.7cm}{B'}_{n+1}=U_{{\rm p}n+1}B_{n+1}U_{{\rm p}n+1}^\dagger\nonumber\\
&{}&=\frac{1}{2}\mbox{diag}(-1,-1,0,0,\ldots,0,0)+\mbox{off-diagonal terms}\nonumber\\
&{}&\hspace{2cm}+a_{n+1}\mbox{diag}(1,-1,1,-1,\ldots,1,-1).
\end{eqnarray}
It should be noted that the off-diagonal terms of Eq. (\ref{bn1}) have been also transformed under the application of $U_{{\rm p}n+1}$. Under this transformation, the following permutations of the matrix elements, in decimal notation, take place. For $l=1,2,\ldots,2^{n-1}$
\begin{equation*}
\begin{array}{l}
(2l-1,2l)\longleftrightarrow(2l-1,2^n+2l-1),\\
 \vspace{0.5cm} (2l,2l-1)\longleftrightarrow(2^n+2l-1,2l-1),\\
\end{array}
\end{equation*}
and, for $l = 2^{n-1}+1,\ldots,2^n$
\begin{equation*}
\begin{array}{l}
(2l-1,2l)\longleftrightarrow(2l-2^n+1,2l),\\
(2l,2l-1)\longleftrightarrow(2l,2l-2^n+1).\\
\end{array}
\end{equation*}
Particularly, the off-diagonal elements of the $2\times 2$ diagonal blocks of ${B}_{n+1}$ are replaced with elements that are $0$. Now, the $2\times 2$ diagonal blocks in ${B'}_{n+1}$ are $-I/2+a_{n+1} Z, a_{n+1}Z, \ldots, a_{n+1}Z$, from the upper left to the lower right. Then we set $a_{n+1}=-1/2$ and apply $U_{{\rm bd}n+1}=\mbox{diag}(I,H,\ldots,H)$ to ${B'}_{n+1}$. All the diagonal blocks but the first one transform to $X$ since $HZH=X$. Then we obtain
\begin{eqnarray}
 &&\hspace{-1.0cm}U_{{\rm bd}n+1}{B'}_{n+1}{U}_{{\rm bd}n+1}^\dagger\notag\\
 &&\hspace{-0.2cm}=\mbox{diag}(-1,0,0,\ldots,0,0)+\mbox{(off-diagonal terms)}.
\end{eqnarray}
Therefore, the unitary transformation ${V'}_{n+1}=U_{{\rm bd}n+1}U_{{\rm p}n+1}(I\otimes {V'}_n)$ and coefficients $b'=b/2$, $\{{a'}_i=a_i/2\}_{i=1}^n$, and $a_{n+1}=-1/2$ satisfy the equality Eq. (\ref{equality}) for all $n\ge 2$ under the imposed condition.

The following example with $n=3$ will clarify the above proof. For $K=\frac{1}{2}(b + a_1 I\otimes I\otimes Z + a_2 I\otimes Z\otimes I) + a_3 Z\otimes I \otimes I$, we have
\begin{eqnarray}
 &{}&\hspace{-1.0cm}(I\otimes {V'}_2) K (I\otimes {V'}_2^\dagger)\notag\\
 &=&\frac{1}{2}I\otimes
\left(\begin{array}{cccc}
-1&0&0&0\\
0&0&\frac{1}{4}e^{-i2\pi/3}&\frac{1}{4}e^{i2\pi/3}\\
0&\frac{1}{4}e^{i2\pi/3}&0&\frac{1}{4}e^{-i2\pi/3}\\
0&\frac{1}{4}e^{-i2\pi/3}&\frac{1}{4}e^{i2\pi/3}&0
\end{array}\right)\notag\\
&{}& \hspace{0.5cm}+a_3 Z\otimes I \otimes I\nonumber\\
&=&\frac{1}{2}\mbox{diag}(-1,0,0,0,-1,0,0,0)+\mbox{off-diagonal terms}\notag\\
&{}& \hspace{0.5cm}+a_3\;\mbox{diag}(1,1,1,1,-1,-1,-1,-1).
\end{eqnarray}
Furthermore, by applying the permutation operator $U_{\rm p3}$, we obtain
\begin{eqnarray}
\label{foradd}
&{}&\hspace{-0.8cm} U_{\rm p3}(I\otimes {V'}_2) K (I\otimes {V'}_2^\dagger) U_{\rm p3}^\dagger\nonumber\\
&{}& =\frac{1}{2}\;\mbox{diag}(-1,-1,0,0,0,0,0,0)+\mbox{off-diagonal terms}\notag\\
&{}&\hspace{1.0cm}+a_3\;\mbox{diag}(1,-1,1,-1,1,-1,1,-1).
\end{eqnarray}
This permutation is realized by two transpositions $(2,5)$ and $(4,7)$ with respect to the row labels (numbered from 1 to 8), and the same transpositions with respect to the column labels. With ket and bra labels, these are transpositions $(\ket{001},\ket{100})$, $(\ket{011},\ket{110})$, $(\bra{001},\bra{100})$, and $(\bra{011},\bra{110})$. Therefore, the off-diagonal terms, that is the second term in the right-hand side of Eq. (\ref{foradd}), are written as the matrix
\begin{widetext}
\begin{equation}
 \fl\frac{1}{2}
\left( \begin{array}{cccc}
\begin{array}{|cc|}
\hline
0&0\\0&0\\\hline
\end{array}&
\begin{array}{cc}
~~~~0~~~~&~~~~0~~~~\\0&0
\end{array}&
\begin{array}{cc}
~~~~0~~~~&~~~~0~~~~\\0&0
\end{array}&
\begin{array}{cc}
~~~~0~~~~&~~~~0~~~~\\0&0
\end{array}\\
\begin{array}{cc}
0&0\\0&0
\end{array}&
\begin{array}{|cc|}
\hline
~~~~0~~~~&~~~~0~~~~\\0&0\\\hline
\end{array}&
\begin{array}{cc}
\frac{1}{4}e^{i2\pi/3} & 0\\ 0 & \frac{1}{4}e^{i2\pi/3}
\end{array}&
\begin{array}{cc}
\frac{1}{4}e^{-i2\pi/3} & 0\\0 & \frac{1}{4}e^{-i2\pi/3}
\end{array}\\
\begin{array}{cc}
0&0\\0&0
\end{array}&
\begin{array}{cc}
\frac{1}{4}e^{-i2\pi/3} & 0\\ 0 & \frac{1}{4}e^{-i2\pi/3}
\end{array}&
\begin{array}{|cc|}
\hline
~~~~0~~~~&~~~~0~~~~\\0&0\\\hline
\end{array}&
\begin{array}{cc}
\frac{1}{4}e^{i2\pi/3} & 0\\0 & \frac{1}{4}e^{i2\pi/3}
\end{array}\\
\begin{array}{cc}
0&0\\0&0
\end{array}&
\begin{array}{cc}
\frac{1}{4}e^{i2\pi/3} & 0\\ 0 & \frac{1}{4}e^{i2\pi/3}
\end{array}&
\begin{array}{cc}
\frac{1}{4}e^{-i2\pi/3} & 0\\0 & \frac{1}{4}e^{-i2\pi/3}
\end{array}&
\begin{array}{|cc|}
\hline
~~~~0~~~~&~~~~0~~~~\\0&0\\\hline
\end{array}
\end{array}\right).
\end{equation}
\end{widetext}
Note that all the off-diagonal elements of the $2\times 2$ diagonal blocks of this matrix have disappeared. Adding up the first and third terms of the right-hand side of Eq. (\ref{foradd}) and substituting $a_3=-1/2$, all the diagonal blocks are mapped to $-Z/2$, except the first diagonal block that remains as $-1/2\left(\begin{array}{cc}2&0\\0&0\end{array}\right)$. Then we use the block-diagonal unitary transformation $U_{\rm bd3}=\mbox{diag}(I,H,H,H)$ where $H$ is the Hadamard operation. Then this leads to
\begin{eqnarray}
&{}&\hspace{-1.0cm}U_{\rm bd3}U_{\rm p3}(I\otimes {V'}_2) K(I\otimes {V'}_2^\dagger) U_{\rm p3}^\dagger U_{\rm bd3}^\dagger\nonumber\\
&{}&\hspace{-0.5cm}=\mbox{diag}(-1,0,0,0,0,0,0,0)+\mbox{(off-diagonal terms)}.
\end{eqnarray}
Thus we found that the unitary transformation ${V'}_3=U_{\rm bd3}U_{\rm p3}(I\otimes {V'}_2)$ satisfies ${V'}_3(\frac{1}{2}b + \frac{1}{2}a_1 I\otimes I\otimes Z + \frac{1}{2}a_2 I\otimes Z\otimes I + a_3 Z\otimes I \otimes I){V'}_3^\dagger=\mbox{diag}(-1,0,0,0,0,0,0,0)+\mbox{(off-diagonal terms)}$. This shows that the equality Eq. (\ref{equality}) holds for $n=3$.

\section{Ancillary qubit for generally witnessing entanglement}
In this section we will show how to detect multipartite entanglement for a general case, without imposing the condition that the state density matrix be diagonal after the disentangling operation \cite{extra}. For this idea, we use a single uninitialized ancillary qubit. Consider a single ancillary qubit initially in a thermal equilibrium  state: $\rho^{[a]}_{\rm in} = p\ket{0}\bra{0}+(1-p)\ket{1}\bra{1}$. For a general proof (gen), consider an entangled state generated from $\ket{0\ldots0}\bra{0\ldots0}$ by using some entangling unitary operation $V$, i.e., $\rho_{\rm gen}=\ket{\psi_{\rm gen}}\bra{\psi_{\rm gen}}=V\ket{0\ldots0}\bra{0\ldots0}V^\dagger$. Then the EW is $W_{\rm gen}=c {\mathbbm{1}}-\rho_{\rm gen}$. For instance for the bipartite case, $W_{\rm gen}$ is non-negative for all positive partial transpose states when $c(\ket{\psi_{\rm gen}})$ is set to the largest Schmidt coefficient of $\ket{\psi_{\rm gen}}$ \cite{GuhneO}. Our interest is not specific to bipartite EWs but applicable to general entanglement.

\begin{figure}[hpt]
\begin{center}
 \includegraphics{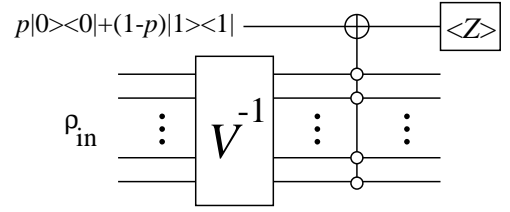}
 \caption{\label{figAncillaWm} Quantum circuit with a polarization measurement to implement the EW $W_{\rm gen}$.}
\end{center}
\end{figure}

In order to test $W_{\rm gen}$ in a single NMR experiment, we introduce the quantum circuit in Fig.\ \ref{figAncillaWm}.
Recall the relation
\begin{equation}
\tr(\rho_{{\rm in}}W_{\rm gen}) =c- \tr \left(V^\dagger\rho_{{\rm in}}V\ket{0\ldots0}\bra{0\ldots0}\right).
\end{equation}
Let us write $\tilde P(0\ldots0)= \tr ( V^\dagger\rho_{\rm{in}}V\ket{0\ldots0}\bra{0\ldots0})$. In the quantum circuit, the diagonal element before the polarization measurement is
\begin{eqnarray}
&{}&\hspace{-1.5cm}\mbox{Diag}\;\{\mbox{C$^n$NOT}[( p\ket{0}\bra{0}\notag\\
&{}&+(1-p)\ket{1}\bra{1}) \otimes(V^\dagger\rho_{\rm{in}}V)]\;\mbox{C$^n$NOT}^\dagger\}.
\end{eqnarray}
Here $\mbox{Diag}\;A$ picks up the diagonal elements from a matrix $A$. After setting $\tilde P(k)=\tr( V^\dagger\rho_{\rm{in}}V\ket{k}\bra{k})$, it is rewritten as
\begin{eqnarray}
&{}&\hspace{-0.8cm}\mbox{C$^n$NOT}\bigl[(p\ket{0}\bra{0}\notag\\
&{}&+(1-p)\ket{1}\bra{1})\otimes \sum_k\tilde P(k)\ket{k}\bra{k} \bigr]\mbox{C$^n$NOT}^\dagger\nonumber\\
&{}&\hspace{-0.8cm}=\tilde P(0\ldots0)[p\ket{1}\bra{1}+(1-p)\ket{0}\bra{0}]\otimes\ket{0\ldots0}\bra{0\ldots0}\notag\\
&{}&+[p\ket{0}\bra{0}+(1-p)\ket{1}\bra{1}]\otimes\sum_{k\not = 0}\tilde P(k)\ket{k}\bra{k}.
\end{eqnarray}
The diagonal part $\mbox{Diag}~\rho^{[a]}_{\rm out}$ of the reduced density operator of the ancillary qubit before the measurement is
\begin{eqnarray}
&{}&\hspace{-0.8cm}\tilde P(0\ldots0)[p\ket{1}\bra{1}+(1-p)\ket{0}\bra{0}]\notag\\
&{}&+[1-\tilde P(0\ldots0)][p\ket{0}\bra{0}+(1-p)\ket{1}\bra{1}]\nonumber\\
&{}&\hspace{-0.8cm}=[\tilde P(0\ldots0)(1-p)+p(1-\tilde P(0\ldots0))]\ket{0}\bra{0}\notag\\
&{}&+[\tilde P(0\ldots0)p+(1-p)(1-\tilde P(0\ldots0))]\ket{1}\bra{1},
\end{eqnarray}
where use has been made of the identity $\sum_k \tilde{P}(k)= 
\tr \rho_{\rm in} = 1$. Thus we have
\begin{equation}
 \tr\rho^{[a]}_{\rm out}Z=(1-2p)[2\tilde P(0\ldots0)-1].
\end{equation}
This leads to
\begin{equation}
 \tilde P(0\ldots0)=\frac{1}{2}-\frac{1}{2(2p-1)}\tr\rho^{[a]}_{\rm out}Z.
\end{equation}
Consequently,
\begin{equation}
 \tr\rho_{\rm{in}}W_{\rm gen}=c-\frac{1}{2}
+\frac{1}{2(2p-1)}\tr\rho^{[a]}_{\rm out}Z.
\end{equation}
Thus, the value of $\tr\rho_{\rm{in}}W_{\rm gen}$ can be found by using the initial polarization $\tr\rho^{[a]}_{\rm in}Z=2p-1$ and the output polarization $\tr\rho^{[a]}_{\rm out}Z$ of the ancillary qubit.

\begin{figure*}
\begin{center}
 \scalebox{0.88}{\includegraphics{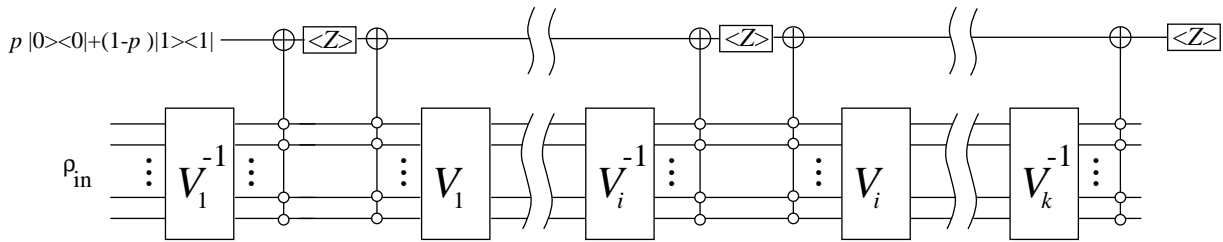}}
\caption{\label{figAncillaWmConcat} Concatenation of the quantum circuit of Fig.\ \ref{figAncillaWm} to test $W_i$'s.} 
\end{center}
\end{figure*}

Extending this method, it is easy to construct a quantum circuit of concatenated EWs to discriminate different types of multipartite entanglement from each other. Suppose we want to measure $k$ types of multipartite entanglement with a set of EWs $\{W_i\}_{i=1}^k=\{c(\ket{\psi_i})-V_i\ket{0\ldots0}\bra{0\ldots0}V_i^\dagger\}_{i=1}^k$, where $\ket{\psi_i}=V_i\ket{0\ldots0}$. Then we can test all of $W_i$ using $k$ measurements as shown in Fig.\ \ref{figAncillaWmConcat}.

For example, tripartite entanglement can be differently written in the form of the GHZ state or the W state. For a general tripartite state $\rho$ this can be checked by applying disentangling operations for the GHZ state (${V'}^{-1}_{\rm GHZ}$) and the W state (${V'}^{-1}_{\rm W}$). This is shown in Fig.~\ref{figAncillaWmConcat} if substitutions are made as $k=2$ and ${V'}^{-1}_1={V'}^{-1}_{\rm GHZ}$ and ${V'}^{-1}_2={V'}^{-1}_{\rm W}$. The entanglement of the state $\rho_{\rm in}$ would be detected by measurements of the ancillary qubit.

In order to realize the quantum circuit Fig.\ \ref{figAncillaWmConcat} with a system such as NMR, one needs to make several free-induction (nondecay) measurements in a single run of a NMR experiment. In a conventional NMR experiment, however, generally one free-induction decay (FID) measurement is made at a particular point where the experiment is ceased. Indeed, there have been experiments involving multiple FID measurements, such as the Cory-48 \cite{cory90} pulse sequence.

Although it is not of particular interest to NMR researchers, in principle it is possible to take an instant free-induction measurement. One possible way is to apply a $\pi/2$ pulse and take signals of the precession for some microseconds duration (that is, small enough to keep coherence), and then apply a $-\pi/2$ pulse. Thus, it is a realistic idea to have multiple free-induction measurements in a single run of an experiment.

\section{Circuit complexity and noise behavior}\label{five}
\subsection{Circuit complexity with diagonal $\rho_{\rm out}$}
\begin{figure}[hpt]
\begin{center}
\includegraphics{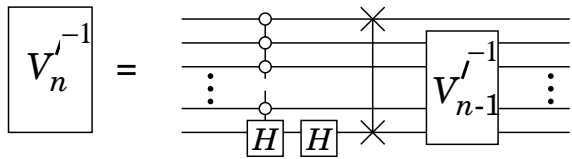}
\caption{\label{figVp}Quantum circuit to compose ${V'}_{n}^\dagger$.}
\end{center}
\end{figure}

The number of two-qubit quantum gates to compose ${V'}_{n}^\dagger$ is $O(n^3)$. This is clear from the quantum circuit of ${V'}_{n}^\dagger$ in Fig.\ \ref{figVp}. Let us write the number of basic gates to compose ${V'}_{n}^\dagger$ by $G(n)$. Then $G(n)=G(n-1)+O(n^2)$, from which we obtain $G(n)=O(n^3)$. Here we used the fact that an in-place C$^nA$ gate ($A$ is a $2\times2$ unitary matrix) is composed of $O(n^2)$ two-qubit gates (see, {\em e.g.}, Ref.\ \cite{NC2000}, p.184).
If we can use a highly selective pulse, the gate C$^nH$ may be performed in a single step, although this usually takes a long time of order $O(2^n)$.

In addition to the circuit complexity to compose ${V'}_{n}^\dagger$, the circuit complexity for the inverse of an entangling operation $V$ should be considered as well. Usually, this circuit complexity is less than $O(n^3)$. For example, a quantum circuit to generate a GHZ-like state from some pure initial state can be composed of several NOT gates, one Hadamard gate, and $n-1$ controled-not (CNOT) gates. Consequently the total circuit complexity is usually $O(n^3)$.

\subsection{Circuit complexity in the method using an ancillary qubit}
In case an ancillary qubit is employed, the dominant circuit complexity is due to the C$^n$NOT gate. On the assumption that the internal circuit of disentangling operations $V^\dagger_\mathrm{X}$ (here, $\mathrm{X}$ may be GHZ, W, etc.) have circuit complexities on the order of $\le n^2$, the total circuit complexity of the quantum circuit of Fig.\ \ref{figAncillaWmConcat} is $O(n^2\times k)$.

\subsection{Noise behavior}
Although the proposed methods enable a single experiment to detect entanglement without copies of states, the size of quantum circuits used in the measurement process is considerably larger than that for usual entangling operations. Thus noise (namely, probabilistic errors) in quantum gates can skew the result of entanglement detection with a higher probability than conventional entanglement detection using multiple copies of a state, assuming that error during preparation of copies of a state is negligible.

In a simple model \cite{D99}, we assume that a quantum gate $U$ acting on the target block $\mathbf{t}$ of qubits suffers from a noise such that a desired unitary transformation takes place with success probability $p_s$; otherwise, a reduced density operator acting on $\mathbf{t}$ becomes a maximally mixed state with failure probability $1-p_s$. The superoperator of this noise is
\begin{eqnarray}
&{}&\hspace{-0.9cm}\mathcal{E}_{p_s,U,\mathbf{t}}(\rho)=p_s(I\otimes U)\rho (I\otimes U^\dagger)\notag\\
&{}&\hspace{-0.6cm}+(1-p_s)(\tr_{\mathbf{t}}\rho)\otimes I^{[i]}/2^{\mathrm{len}(\mathbf{t})},
\end{eqnarray}
where $\mathrm{len}(\mathbf{t})$ is the number of qubits in $\mathbf{t}$; $I^{[i]}/2^{\mathrm{len}(\mathbf{t})}$ is the maximally mixed state of $\mathbf{t}$ and $I$ in the first term is the identity matrix acting on the rest of the qubits.

\begin{figure}[hpt]
\begin{center}
 \scalebox{0.5}{\includegraphics{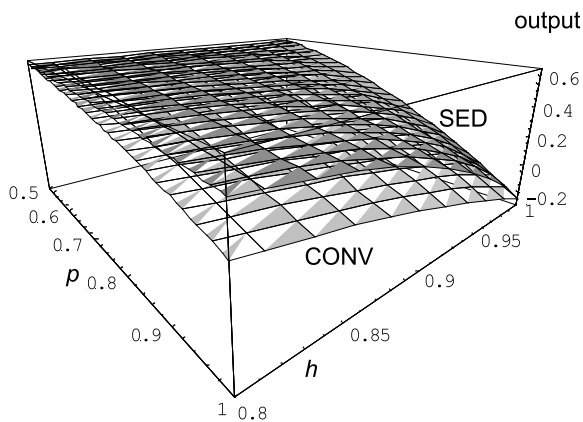}}
 \caption{\label{EWComparison} Output of conventional entanglement witness (CONV) and that of single-experiment-detectable entanglement witness (SED) without ancillary qubit as functions of the probability $p$ of starting states and the success probability $h$ of quantum gates.}
\end{center}
\end{figure}

Although this noise model is rather simple, analytical calculation of the output of a quantum circuit of our interest under this noise is quite complicated and does not result in a tractable equation. Instead, we present a result for a particular example of the quantum circuit shown in Fig.\ \ref{figone} in the case of a three-qubit input starting from the thermal state $[p|0\rangle\langle0|+(1-p)|1\rangle\langle1|]^{\otimes 3}$. We set the success probability as $p_s=h$ for single-qubit gates and $p_s=h^2$ for CNOT gates. Noise is assumed to exist in individual quantum gates including the entangling operation $V$. We investigate the output $\tr(\rho_{\rm in} W)$ in Eq. (\ref{equality}) under the above noise. We choose the conventional entanglement witness for the class of the GHZ state given in Eq. (\ref{convEWGHZ}) and decomposition into SED EW using the method that we have introduced in Sec.\ 3. Numerical results of outputs of entanglement witnesses are plotted against $p$ and $h$ in Fig.~\ref{EWComparison}.
As illustrated in the figure, positive values are returned in the range of $p$ in which the noise-free entanglement witness (i.e., the case of $h=1$) returns negative values. A SED EW is more fragile against noise than a conventional one in the sense that the range of $h$ in which negative values are returned is small. Nevertheless, it is important that we never obtain a negative value for a separable state even under noise in this example. For a general case, a mathematical proof for returning non-negative values for all separable states is not easy because it is strongly dependent on the structure of the quantum circuit that has been used for the scheme introduced in this paper.

One way to improve the robustness is to increase the value of $p_s$. Recently, a high gate fidelity was reported by using a classical numerical optimization in NMR quantum computing ({\em e.g.} Ref.\ \cite{SH05}) and this technique may be applicable for this purpose. This technique can also reduce the number of pulses needed to implement a large quantum gate. Thus it seems a possible way to make proposed entanglement detection methods practical in the future.

\section{Summary}
We proposed two schemes to reconstruct an entanglement witness into nonlocal operations and local measurements so that a single experiment without copies of a state is sufficient. In one scheme, an ancillary qubit is not required but the quantum state must satisfy some condition, while an uninitialized ancillary qubit is required in the other scheme where no condition is imposed on the quantum state. Computational complexities and noise behavior have been discussed.

\begin{acknowledgments}
We would like to thank Masato Koashi for pointing out an error in the description of concatenated EWs in the draft and Kazuyuki Takeda for discussions. R.R. is grateful to Vlatko Vedral for helpful discussions. A.S. is supported by the JSPS. M.N. would like to thank MEXT for partial support (Grant No. 13135215). M.K. is supported by CREST of Japan Science and Technology Agency.
\end{acknowledgments}

\end{document}